\documentclass[english,aps,preprint,nofootinbib]{revtex4-2}%
\usepackage[T1]{fontenc}
\usepackage[latin9]{inputenc}
\usepackage{amsmath}
\usepackage{amssymb}
\usepackage{babel}
\usepackage{amsfonts}
\usepackage{graphicx}
\usepackage{hyperref}
\usepackage{xcolor}%
\begin{document}

\title{Making sense of relativistic Wigner friend scenarios: a problem for unitary
accounts of quantum measurements ?}
\author{J. Allam}

\affiliation{Laboratoire de Physique Th\'eorique et Mod\'elisation, CNRS Unit\'e 8089, CY
	Cergy Paris Universit\'e, 95302 Cergy-Pontoise cedex, France}
\author{ A. Matzkin}
\affiliation{Laboratoire de Physique Th\'eorique et Mod\'elisation, CNRS Unit\'e 8089, CY
	Cergy Paris Universit\'e, 95302 Cergy-Pontoise cedex, France}

\begin{abstract}
Wigner-friend scenarios -- in which external agents describe a closed
laboratory containing a friend making a measurement -- highlight the
difficulties inherent to quantum theory when accounting for measurements. In
non-relativistic scenarios, the difficulty is to accommodate unitary evolution
for a closed system with a definite outcome obtained by the friend. In
relativistic scenarios the tensions between quantum theory and relativity
induce additional constraints. A generic property of relativistic scenarios is
the frame-dependence of state update upon a measurement. Based on a definite
example, we will show that this property leads to inconsistent accounts for
outcomes obtained in different reference frames. We will further argue that these results point to some
fundamental inadequacy when attempting to model actions taken by a complex
agent as unitary operations made on simple wavefunctions.

\end{abstract}
\maketitle

\newpage

\bigskip


\section{Introduction}

Wigner friend scenarios bring to forth the tension existing between unitary
evolution, that according to standard quantum theory applies to closed
systems, and the projection postulate, that applies when an agent performs a
measurement and observes a result. In his seminal paper \cite{wigner}, Wigner
had already noted the ambiguity of an external observer's description of a
perfectly isolated laboratory containing an agent making a measurement.
Indeed, the friend having measured the spin of a particle cannot remain in a
state of "\emph{suspended animation}" until the external observer asks her
what outcome she obtained. Wigner suggested in Ref. \cite{wigner} that the
Schrödinger equation needs to be supplemented with a non-linear term when a
measurement takes places, in order for the state update to be the same for the
friend and the external observer.

However, standard quantum mechanics has no provision for an additional
non-linear term and remains ambiguous on how the state should be updated after
a measurement took place. Should a closed laboratory be described as no
differently from any isolated quantum system, even if it contains an agent
(the friend)? In this case, for an external observer labeled W, the quantum
state of the laboratory evolves unitarily, despite the fact that the friend
obtained a definite outcome. Even assuming W endorses a unitary description,
there is an additional ambiguity as the friend's outcome may be assumed to be
unique and objectively defined for anyone \cite{renner}, or instead constitute
a fact for the friend only, this fact not being defined for external observers
such as W who do not apply the projection postulate \cite{brukner,cai}.

\begin{figure}[ptb]
\centering
\includegraphics[width=0.5\textwidth]{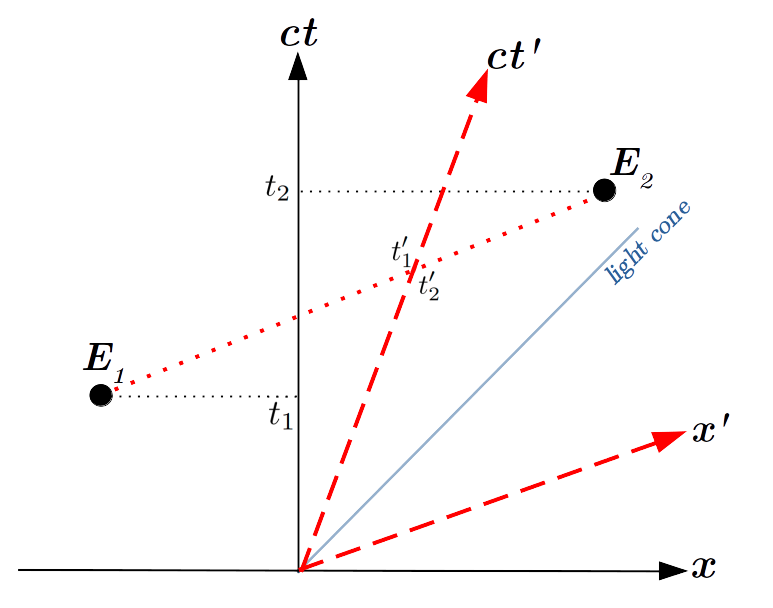}\caption{Space-like separated
measurements of a bipartite entangled state in two reference frames. The
measurement event E$_{1}$ takes place before E$_{2} $ in the frame ($x$,
$ct$), while they are simultaneous in the frame ($x^{\prime}$, $ct^{\prime}$).
During intermediate times ($t_{1}<t<t_{2}$), the description of the quantum
state is different in frame ($x$, $ct$) in which a measurement took place,
from the description in the reference ($x^{\prime}$, $ct^{\prime}$) in which
the state remains entangled. }%
\label{fig:photo}%
\end{figure}

In this context, relativistic considerations can be introduced in Wigner
friend scenarios essentially in order to examine the constraints arising from
different time-orderings of space-like separated events.\ Indeed, if event
$E_{1}$ precedes a space-like separated event $E_{2}$ in a given reference
frame, there are inertial reference frames in which $E_{2}$ precedes $E_{1}$,
see Fig. \ref{fig:photo}. This leads to well-known consequences concerning
state-update \cite{AA2928,peres}. For example if Alice and Bob share two spin
1/2 particles in an entangled state, say
\begin{equation}
\alpha\left\vert +u\right\rangle \left\vert -v\right\rangle +\beta\left\vert
-u\right\rangle \left\vert +v\right\rangle +\gamma\left\vert -u\right\rangle
\left\vert -v\right\rangle \label{ex1}%
\end{equation}
where $u$ (resp. $v$) is the direction chosen by Alice (resp.\ Bob) to measure
the spin component, then in a reference frame in which Alice measures first,
the state after her measurement is updated to $\left\vert -v\right\rangle $
(if Alice obtained $+1$) or to $\beta\left\vert +v\right\rangle +\gamma
\left\vert -v\right\rangle $ (if she obtained $-1$). If Alice and Bob's
measurements are space-like separated events, there is an inertial frame in
which Bob's measurement happens first, and hence the state is updated either
to $\left\vert -u\right\rangle $ or to $\alpha\left\vert +u\right\rangle
+\gamma\left\vert -u\right\rangle $. The consensus \cite{peres-review} is that
the frame dependence of the quantum states at intermediate times is not a
problem, given that the outcomes and probabilities are identical in both
reference frames (or more generally, they are related by a Lorentz transform).
But in situations in which an outcome might not be associated with a state
update (in our case, the friend's measurement), it becomes necessary to
analyze the implications of dealing with quantum states that might describe
different physics in distinct inertial reference frames.

In the original thought experiment devised by Wigner \cite{wigner}, it is
straightforward to compute different probabilities for the external agent's
outcome depending on whether this agent applies a state update after the
friend's measurement (on the ground that the projection postulate should
apply), or a unitary evolution (on the ground that the Schrödinger equation
should be used since the laboratory is an isolated system). In all cases, the
friend does update her state after completing her measurement.\ This gives
rise, if unitary evolution is assumed, to a contradiction when considering two
friends (each friend sitting in an isolated laboratory) sharing an entangled
state (see the review \cite{renner} and Refs. therein, as well as
\cite{lombardi,relano,jordan,WFWM,toys,histories,wfr,baumann}).
Technically, the contradiction is at the level of joint probabilities for
measurement outcomes, given that a unitary account involves interferences,
implying that such joint probabilities cannot be obtained as a marginal
distribution -- a well-known situation in quantum mechanics (e.g.
non-contextual inequalities, or Bell-type inequalities).

In a Wigner friend scenario in a relativistic setting, an additional
ingredient comes into play: if a state of the type given by Eq. (\ref{ex1})
describes a Wigner friend setup, one might question whether the fact that the
intermediate state after a measurement on an entangled pair is different in
two inertial reference frames might not lead to a new type of contradiction if
unitary evolution is assumed. Drawing on a recent work \cite{us}, we wish to
examine in this paper the implications of what appears to be a generic
property of relativistic Wigner friend scenarios: a frame-dependence of the
outcomes observed by an external observer. We will base our discussion on a
definite example to be described in Sec.\ \ref{an} after having briefly
recalled the salient features characterizing Wigner friend scenarios and the
role of relativistic constraints (Sec. \ref{wfs}). We will then analyze and
discuss in Sec.\ \ref{disc} the main implications of relativistic models; we will argue
in particular that describing a decision-making agent (the friend) by a
quantum state evolving unitarily could be the underlying issue.

\section{Wigner friend scenarios\label{wfs}}

In the original Wigner Friend scenario (WFS) Wigner \cite{wigner} introduces a
sealed laboratory L in which a friend F performs a Stern-Gerlach experiment on
an atomic spin, while an agent W is outside the isolated laboratory and
ultimately measures the quantum state of the laboratory (in the same basis) by
asking F what she obtained. The spin is initially in state
\begin{equation}
\left\vert \psi(t_{0})\right\rangle =\alpha\left\vert +\right\rangle
+\beta\left\vert -\right\rangle \label{s00}%
\end{equation}
and the isolated lab is assumed to be described by the quantum state%
\begin{equation}
\left\vert L(t_{0})\right\rangle =\left\vert \psi(t_{0})\right\rangle
\left\vert m_{0}\right\rangle \left\vert \varepsilon_{0}\right\rangle
\end{equation}
where $\left\vert m_{0}\right\rangle $ and $\left\vert \varepsilon
_{0}\right\rangle $ denote the initial states of the pointer and environment respectively.

The issue is whether a sealed laboratory with an observer inside should be
described, by an external agent as evolving unitarily (on the ground that the
laboratory is an isolated closed quantum system) or as a statistical mixture
(assuming that any measurement implies a state update\footnote{In Ref.
\cite{wigner}, Wigner supports the idea that the Schrödinger equation must be
supplemented with nonlinear terms for conscious agents, though he will later
change his mind \cite{history,ballentine}.} for all observers). In principle
(assuming that the laboratory is still a quantum object after F's
measurement), W can measure the laboratory L in a basis different from F's
measurement. Indeed if $\left\vert \psi(t_{0})\right\rangle $ is measured in
the $\left\vert \pm\right\rangle $ basis, unitary evolution leads to
\begin{equation}
\left\vert L(t)\right\rangle =\alpha\left\vert L_{+}(t)\right\rangle
+\beta\left\vert L_{-}(t)\right\rangle \label{s01}%
\end{equation}
where
\begin{equation}
\left\vert L_{\pm}\right\rangle \equiv\left\vert \pm\right\rangle \left\vert
m_{\pm}\right\rangle \left\vert \varepsilon_{\pm}\right\rangle \label{s01s}%
\end{equation}
are the states of the laboratory having inherited the $\pm$ spin outcome; W
can choose to measure L in any basis spanned by $\left\{  \left\vert
L_{+}\right\rangle ,\left\vert L_{-}\right\rangle \right\}  $. However such a
measurement will then destroy F's measurement records (and any memory of the
record) \cite{EPL}, so that no conflicting statements between the internal and
external observers can be obtained \footnote{Of course, the computed
probabilities for W are different according to whether unitary evolution or
the projection postulate are applied.}.

Extended scenarios, introduced in Ref.\ \cite{deutsch}, build on the WFS by
combining more observers inside or outside isolated laboratories in order to
formulate stronger assumptions leading to a consistent description of the
agents' observations. For example F can open a communication channel by which
she informs W that she obtained a definite outcome (without revealing this
outcome) \cite{deutsch}, somewhat circumventing the destruction of F's records
by W's measurement. Indeed, the joint existence of F's and W's observations
remains problematic even when the postulates for which a WFS makes sense are
endorsed. Additional assumptions concerning the validity of inferences made by
the agents using the theory need to be made \cite{renner}. Alternatively, the
existence of joint facts can be denied in favor of observer-dependent facts
\cite{brukner}. Note it can also be argued that claiming F made an observation
while keeping interference terms between the branches corresponding to the
friends possible outcomes is contradictory, as it violates the uncertainty
principle \cite{soko-matz-entropy}, but then this implies unitary evolution
should be discarded when a measurement takes place.

\section{State update in relativistic Wigner friend scenarios\label{an}}

\subsection{Relativistic constraints}

An additional ingredient appearing in a relativistic setting is
that for space-like separated events, the time-ordering valid in one reference
frame can be reversed in another inertial reference frame (see Fig.
\ref{fig:photo} for an illustration). This is the only specific relativistic
ingredient we will be using, as the dynamical laws do not play
any role in WFS, and we can safely assume situations in which energies and
velocities remain sufficiently small to avoid pair creation or the frame
dependence of spin. We will further assume that the state update of a
separated multipartite quantum system takes place instantaneously in any
inertial reference frame -- an assumption that is more or less standard, if
not consensual \cite{peres,peres-review,fayngold} . This implies that the
description of the quantum state at intermediate times between preparation and
final measurements will not be the same in different reference frames -- these
intermediate states are actually unrelated by any Lorentz transform, such as
in the example given below Eq. (\ref{ex1}). This is a generic situation in
relativistic settings and is well-known -- at least when the states describe
unambiguous quantum systems -- to have no observational consequence, given
that the final measurement outcomes and their probabilities remain the same
(or are related by a Lorentz transformation) in all reference frames
\cite{peres,AArel}. However, in a WFS, the friend is an agent who is described
by a quantum state. We now put forward a model in which the friend's actions
depend on this intermediate quantum state, leading to an inconsistent
description in different reference frames.

\begin{figure}[ptb]
	\centering
	\includegraphics[width=0.8\textwidth]{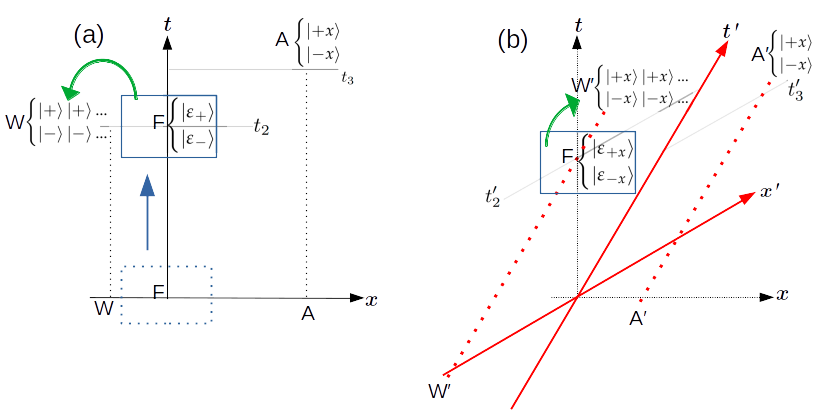}\caption{Schematic representation
		of the protocol detailed in Sec. \ref{wus}, in the reference frames $\mathcal{R}$ [(a)]
		and  $\mathcal{R'}$  [(b)]. The sealed laboratory with the friend F inside 
		is represented by the box at the time ($t_2$ in $\mathcal{R}$ and $t^{\prime}{_2}$ in $\mathcal{R'}$) F sends qubits prepared in a state according to her observation, as determined by the environment states ($|\varepsilon_\pm \rangle$  in $\mathcal{R}$ and $|\varepsilon_{\pm x}\rangle$ in $\mathcal{R'}$). }%
	\label{fig-scheme}%
\end{figure}

\subsection{State update and agent's actions\label{wus}}

Let us start with two particles created in the entangled state
\begin{equation}
|\psi_{FA}\rangle=1/\sqrt{2}\big(|+\rangle_{F}|+\rangle_{A}+|-\rangle
_{F}|-\rangle_{A}\big) \label{fa1}%
\end{equation}
that are sent to the agents F (friend) and A (Alice). A and F will measure their
respective particle's spin at space-like separated events. F is situated
inside an isolated laboratory; next to this lab there is an external observer
W equipped with several spin-measuring devices (see Fig. \ref{fig-scheme} for a schematic representation). The total initial state is
therefore
\begin{equation}
|\Psi(t_{0})\rangle=|\psi_{FA}\rangle\left\vert m_{0}\right\rangle
|\varepsilon_{0}\rangle\label{fa2}%
\end{equation}
where $\left\vert m_{0}\right\rangle $ and $|\varepsilon_{0}\rangle$ are the
states of F's measuring apparatus and the lab's environment respectively. The
lab is described by the compound state $|L_{\pm}\rangle=|\pm\rangle_{F}%
|m_{\pm}\rangle|\varepsilon_{\pm}\rangle$, where $|\pm\rangle_{F}$ represents
F's spin state, $|m_{\pm}\rangle$ represents F's measuring device, and
$|\varepsilon_{\pm}\rangle$ represents the environment state inside the lab.

Let us first describe the protocol in a reference frame $\mathcal{R}$ in which the laboratory is at rest.

\begin{itemize}
\item At time $t_{1}$, F measures her spin in the $|\pm\rangle$ basis, and
assuming unitarity, the spin superposition is inherited by the state of the
laboratory, i.e. the apparatus and environment states of Eq. (\ref{fa2}) are
transformed to mutually orthogonal states $\left\vert m_{\pm}\right\rangle
\left\vert \varepsilon_{\pm}\right\rangle $ after coupling with the spins, so
that according to an external observer
\begin{equation}
|\Psi(t_{1})\rangle=\frac{1}{\sqrt{2}}\big(|L_{+}\rangle|+\rangle_{A}%
+|L_{-}\rangle|-\rangle_{A}\big)
\end{equation}
where $|L_{\pm}\rangle=|\pm\rangle_{F}|m_{\pm}\rangle|\varepsilon_{\pm}%
\rangle$ is the state of the laboratory correlated with a $\pm1$ outcome for
F's measurement.

\item Following her measurement, F immediately resets the states of the spin
and the measuring device to a pre-assigned state $|s_{0}\rangle_{F}%
|m_{0}\rangle$. By defining $|\tilde{L}_{\pm}\rangle=|s_{0}\rangle_{F}%
|m_{0}\rangle|\varepsilon_{\pm}\rangle$, the state can now be written at
$t=t_{1}^{\ast}>t_{1}$ as
\begin{equation}
|\Psi(t_{1}^{\ast})\rangle=\frac{1}{\sqrt{2}}\big(|\tilde{L}_{+}%
\rangle|+\rangle_{A}+|\tilde{L}_{-}\rangle|-\rangle_{A}\big). \label{fa4}%
\end{equation}

\item At time $t_{2}$, F declares the outcome she observes, which depends on
the environment states $|\varepsilon_{\pm}\rangle\ $(as any record of the
outcome must be part of the environment). She does so by creating many
particles in the state $|+\rangle$ (if she observes $+$) or $|-\rangle$ (if
she observes $-$). She sends the particles she has just generated outside the
lab to W via a quantum communication channel so that $|\Psi(t_{2})\rangle$
becomes after the particles creation%
\begin{equation}
\frac{1}{\sqrt{2}}\big(|\tilde{L}_{+}\rangle|+\rangle_{A}\left\vert
+\right\rangle _{W}\left\vert +\right\rangle _{W}...+|\tilde{L}_{-}%
\rangle|-\rangle_{A}\left\vert -\right\rangle _{W}\left\vert -\right\rangle
_{W}...\big), \label{fa5}%
\end{equation}
where the index $W$ labels the particles sent to W. W must measure at least
one particle in the $z$ basis, and one particle in the $x$ basis; at this
point, the state given by Eq. (\ref{fa5}) is updated to the $+$ or the $-$
branches, and W can conclude that F has observed $+$ or $-$, i.e. an outcome
of a spin-z measurement.

\item At time $t_{3}$, Alice performs a spin measurement on her particle in
the $|\pm x\rangle$ basis, which does not affect the final result obtained by W.
\end{itemize}

In another reference frame $\mathcal{R}^{\prime}$ in motion relative to
$\mathcal{R}$, the instant $t_{3}^{\prime}$, at which Alice measures in the
$|\pm x\rangle$ basis takes place before $t_{2}^{\prime}$, the time at which F
creates particles according to the state she observes. In $\mathcal{R}%
^{\prime}$ we can think of A$^{\prime}$ and W$^{\prime}$ as moving observers
relative to the particle and the laboratory, who are synchronized to pass by
the particle and laboratory at the times of measurements, $t_{3}^{\prime}$ and
$t_{2}^{\prime}$ respectively. Hence A$^{\prime}$ measures the system in state
given by Eq. (\ref{fa4}) and after A$^{\prime}$'s measurement, state update
leads to%
\begin{equation}%
\begin{cases}
|\Psi_{+x}^{\prime}(t_{3}^{{\prime}})\rangle=\frac{1}{\sqrt{2}}\left(
|\tilde{L}_{+}\rangle+|\tilde{L}_{-}\rangle\right)  \left\vert +x\right\rangle
_{A}\equiv|\tilde{L}_{+x}\rangle\left\vert +x\right\rangle _{A}\\
|\Psi_{-x}^{\prime}(t_{3}^{{\prime}})\rangle=\frac{1}{\sqrt{2}}\left(
|\tilde{L}_{+}\rangle-|\tilde{L}_{-}\rangle\right)  \left\vert -x\right\rangle
_{A}\equiv|\tilde{L}_{-x}\rangle\left\vert -x\right\rangle _{A}%
\end{cases}
\end{equation}
depending on whether A's outcome is $+x$ or $-x$. We have used $|\tilde
{L}_{\pm x}\rangle=|s_{0}\rangle_{F}|m_{0}\rangle|\varepsilon_{\pm x}\rangle$
and defined the superposition of environment states as $\big(|\varepsilon
_{+}\rangle\pm|\varepsilon_{-}\rangle\big)/\sqrt{2}=|\varepsilon_{\pm
x}\rangle$. Although it can be argued \cite{us} that the states $|\varepsilon
_{\pm x}\rangle$ correspond to environmental states after an outcome $\pm x$
has been observed, it is enough for the sake of the argument presented here to
notice that the states $|\varepsilon_{\pm x}\rangle$ represent states that are
distinct from those described by either $|\varepsilon_{+}\rangle$ or
$|\varepsilon_{-}\rangle$. Since F's observation relies on records that are
part of the environment states, F now observes either $+x$ or $-x$. She follows
the protocol and sends qubits to W$^{\prime}$ in the state $\left\vert
+x\right\rangle $ or $\left\vert -x\right\rangle $ depending on whether she
has observed $+x$ or $-x$. Therefore at $t_{2}^{\prime}$, after F creates the
qubits, the quantum state in $\mathcal{R}^{\prime}$ is either of the two
following states:%
\begin{equation}%
\begin{cases}
|\tilde{L}_{+x}\rangle|+x\rangle_{A}\left\vert +x\right\rangle _{W}\left\vert
+x\right\rangle _{W}...\\
|\tilde{L}_{-x}\rangle|-x\rangle_{A}\left\vert -x\right\rangle _{W}\left\vert
-x\right\rangle _{W}...
\end{cases}
.
\end{equation}
Again W$^{\prime}$ makes at least one measurement in the $z$ basis and one in
the $x$ basis, and we assume he has received enough qubits in order to
characterize F's outcome, namely $+x$ or $-x$.

Therefore in $\mathcal{R}^{\prime}$, the external observer receives qubits in
different states than those received in $\mathcal{R}$. This contradiction is a
consequence of (i) the intermediate state being different in distinct
reference frames (a generic property of relativistic setting as we have
mentioned above) and (ii) the fact that F is an agent whose action depends
indirectly on this intermediate quantum state (through the state of the
environment inside the laboratory).

\section{Discussion\label{disc}}

\subsection{Inconsistencies in non-relativistic scenarios}

In the usual non-relativistic Wigner friend scenarios, the tension is between
the application of the projection postulate after a measurement, and a unitary
description for a closed system. Mixing both type of evolutions (assuming
measurement outcomes exist objectively for any observer in conjunction with a
unitary description by the external observers) leads to inconsistencies. At
the formal level, this is strictly equivalent \cite{soko-matz-entropy} to a
double-slit experiment in which one would know which slit the particle took
(this would correspond to a friend's measurement) while still maintaining an
interference pattern at the screen (this is the description employed by the
external observers measuring measuring a friend's laboratory). Of course, at
the interpretational level such an analogy is unwarranted, given that the
issue at stake is to characterize a measurement taking place inside a system
that would be described unambiguously as following unitary evolution if an
agent was not placed inside. Requesting the existence of all the agents'
measurement records is mathematically equivalent to assuming the existence of
a joint probability distribution for the outcomes of each friend and its
corresponding Wigner \cite{persistent,kastner}. Quantum theory has no
provision for the existence of joint probability distributions for
incompatible observables \cite{lombardi} -- there is indeed no common
eigenbasis and by measuring the lab Wigner's measurement destroys
his corresponding friend's record \cite{EPL}.\ 

The alternative to unitary evolution is to apply state update after any
measurement, irrespective of whether the measurement takes place in a closed
system. Note that we are not interested here in giving a mechanism accounting for
state update \footnote{Such as e.g. what would be proposed by objective
collapse theories, or through effective collapse in some versions of the de
Broglie-Bohm interpretation that discard for all practical purposes\ the action of the empty
waves after a measurement.} -- after all this is nothing but the projection
postulate as given in textbook quantum mechanics. Note also that conceptually,
whether state update should apply to all agents should not be conflated with
the hypothesis concerning the application of quantum mechanics to isolated
macroscopic laboratories containing an agent -- if the latter hypothesis is
not fulfilled, there are no Wigner friend scenarios, but one can still endorse
this hypothesis and reject a unitary description. Doing so leads to the
conundrum encapsulated in Bell's question -- \textquotedblleft\textit{ What exactly
qualifies some physical systems to play the role of
'measurer'?}\textquotedblright\ \cite{bell}.

\subsection{Role of state update in relativistic contexts}

In addition to the inconsistencies mentioned above, relativistic Wigner friend
scenarios have to deal with the consequences of frame-dependent intermediate
states. {Such states are not Lorentz transforms of one another. If the quantum
state refers to particles, as in Eq. (\ref{ex1}) it is known that having
different intermediate states in distinct inertial frames does not lead to any
observational consequence \cite{peres}.\ Actually it turns out it is
impossible to characterize an intermediate state by itself \cite{tobepub}.
However in Wigner friend scenarios, the quantum state describes an agent and
its environment. By definition an agent acts, and the operations the agent
undertakes can only depend on the quantum state of the laboratory (as long as
the laboratory is isolated). This is also the case in non-relativistic
versions, where laboratories are prone to superposition, interference and
state update.}

{Inconsistencies arise in a relativistic setting because the intermediate
updated state is different in each reference frame. Then the friend's
operations become frame-dependent, and this can be inferred by an external
observer making interventions on the laboratory. In the model of
Sec.\ \ref{wus}, in the frame $\mathcal{R}$ the environment is in one of the
states $|\varepsilon_{\pm}\rangle$ and the friend prepares accordingly qubits
in the states $|\pm\rangle,$ whereas in $\mathcal{R}^{\prime}$ the state of
the laboratories is updated after A}$^{\prime}$'s measurement.\ {This leaves
the environment in states $|\varepsilon_{\pm x}\rangle$ and the friend now
prepares and sends to W qubits prepared in the }$x$ basis, {in either of the
states $|\pm x\rangle$. Hence the inconsistency between the outcomes predicted
in different reference frames is due to state update in contexts in which the
ordering of space-like events depends on the observers' reference frame.} Note that if the
external observer describes the friend's measurement by applying the
projection rule, rather than unitary evolution, then no contradiction between
outcomes in different frames is obtained.

\subsection{Unitary agents}

The inconsistent descriptions of Wigner friend scenarios in two different
reference frames is of course highly problematic. The problem arises when
combining in a relativistic context the asumptions generally endorsed in
Wigner friend scenarios with the instantaneous state update rule of standard
quantum mechanics. The validity of the state update rule in accounting for
quantum correlations has never been questioned.\ However, the instantaneous
character of state update conflicts with the quantum state representation of
an agent: this can potentially lead to signaling. 

Indeed, consider the entangled state given by Eq. (\ref{fa1}) in which the
friend measures her qubit in a sealed laboratory. Then after the friend's
measurement (there is no state update) A measures the $\theta$ component of her spin, so that we can
write%
\begin{align}
|\Psi(t_{1})\rangle & =\frac{1}{\sqrt{2}}\left(  \cos\frac{\theta}{2}%
|L_{+}\rangle+\sin\frac{\theta}{2}|L_{-}\rangle\right)  |+\theta\rangle
_{A}\label{ee1}\\
& +\frac{1}{\sqrt{2}}\left(  -\sin\frac{\theta}{2}|L_{+}\rangle+\cos
\frac{\theta}{2}|L_{-}\rangle\right)  |-\theta\rangle_{A}.\label{ee2}%
\end{align}
After A's measurement, the state of the laboratory updates instantaneously to
one of the expressions between $(...)$ in Eqs. (\ref{ee1}) or (\ref{ee2}),
which are here mutually orthogonal. Now any operation by which the agent knows
about the state of the laboratory gives rise to signaling; for instance the
friend could determine instantaneously the measurement direction $\theta$
chosen by A (a statement that does not make sense from a relativistic
standpoint). We stress here that we allow the friend to perform operations
that might not necessarily be represented by unitary transformations, on the
ground that since the friend is an agent, then she can undertake any
operations that an agent outside the sealed lab could perform. For instance
copying a known qubit in an arbitrary state {is impossible to implement
unitarily; doing so would would break the linearity of quantum mechanics and
lead to signaling \cite{ghirardi2}. Note further that as emphasized by Peres
}\cite{peres}, an instantaneous update of the quantum state should be tied to
a mathematical computation rather than  to a physical process to avoid conflicting
with relativistic constraints.

{This situation leads to fundamental questions concerning the validity of modeling an agent
with simple quantum states obeying standard (linear) quantum mechanics. This
question has not received much attention up to now in the context of Wigner
friend scenarios, but the operations available to an agent described by
unitary quantum mechanics are restricted when compared to what one would
expect from a classical agent. Put differently, describing a complex
decision-making agent by simple wavefunctions evolving unitarily might turn
out to be an oversimplification leading to inconsistencies. This brings us
back to the well-known difficulties in coping with the measurement problem --
a theory whose dynamics is based on wavefunctions evolving unitarily cannot
account for single outcomes \cite{maudlin}, although this is precisely an
assumption that is frequently made when studying Wigner friend scenarios based
on the \textquotedblleft closed system\textquotedblright\ property of the
sealed laboratory.}

\section{Conclusion\label{conc}}

We have analyzed relativistic constraints that appear in Wigner friend
scenarios. While in non-relativistic Wigner friend scenarios inconsistencies
arise when attempting to describe probabilities for joint outcomes of
incompatible observables, an additional property generic to relativistic
scenarios {is that instantaneous state update on entangled states leads to an
intermediate quantum state that is specific to a given reference frame. We
have seen in the example given in Sec. \ref{wus} that this leads to
inconsistent outcomes in different reference frames, depending on whether the
update took place before or after the external agent's measurement. We have
further argued that relativistic constraints bring to the fore the possible
inadequacy of accounting for decision-making agents by describing them with
simple quantum states evolving unitarily. }

\end{document}